\documentclass[12pt]{article}

\usepackage{amsfonts,amsmath,amssymb,amsthm}

\usepackage{cite}

\begin{document}

\begin{center}
{\bf \Large Conservation laws of the system of equations\\ of
one-dimensional shallow water over uneven bottom in Lagrange's
variables}
\end{center}

\centerline{\large Alexander~V. Aksenov$^{a,b}$ and Konstantin~P. Druzhkov$^{a,b}$}

\begin{center}
$^{a}$ Lomonosov Moscow State University, 1 Leninskiye Gory,\\
Main Building, 119991 Moscow, Russia\\
\end{center}

\begin{center}
$^{b}$ Keldysh Institute of Applied Mathematics RAS,\\
4~Miusskaya~Square, 125047 Moscow, Russia
\end{center}

\centerline{E-mail: aksenov.av@gmail.com and Konstantin.Druzhkov@gmail.com}

\

\centerline{\textbf{Abstract}}

The system of equations of one-dimensional shallow water over
uneven bottom in Euler's and Lagrange's variables is considered.
Intermediate system of equations is introduced. Hydrodynamic
conservation laws of intermediate system of equations is used to find
all first order conservation laws of shallow water equations in
Lagrange's variable for all bottom profiles. The obtained conservation
laws are compared with the hydrodynamic conservation laws of the system
of equations of one-dimensional shallow water over uneven bottom in
Euler's variables. Bottom profiles are given for which there are
additional conservation laws.

\


\section{Introduction}

There are various approaches to finding conservation laws of equations
of mathematical
physics~\cite{Noether,Olver,VinKr,BlumanChAnco,Ibr,BlChA}. The most
widely known method of constructing of conservation laws is based on
Noether's theorem~\cite{Noether}. This method uses symmetries.

Many works are devoted to the construction of conservation laws of
equations in hydro- and
gas-dynamics~\cite{Shmygl,MelSK,MelS,PolZ,PolZZh}.

The hydrodynamic conservation laws of the one-dimensional shallow water
equations over uneven bottom in Euler's variables were obtained
in~\cite{AksDr}.

In the present work, the first-order conservation laws of the shallow
water equations in Lagrangian's variables for all bottom profiles are
obtained without using of symmetries.

\section{Basic Equations}

In dimensionless variables, the system of one-dimensional shallow-water
equations over an uneven bottom has the following form~\cite{Stoker}:
\begin{equation}
\begin{aligned}
&u_t + uu_x + \eta_x = 0,\\
&\eta_t + ((\eta + h(x))u)_x = 0.
\label{SW}
\end{aligned}
\end{equation}
Here $h(x)$ is the thickness of the unperturbed layer of the liquid,
$u=u(x,t)$ is the depth-average horizontal velocity, $\eta=\eta(x,t)$
is the deviation of the free surface $(\eta(x, t) + h(x)\ge 0)$. The
bottom profile is given by the relation $z=-h(x)$ ($z$ is the vertical
coordinate).

\textbf{Remark~1.} The system of equations~\eqref{SW} is similar to the
system of equations of one-dimensional gas
dynamics~\cite{Chernyi,CourFriedr}.

Using the second equation of the system of equations~\eqref{SW}, we
introduce a new variable $m=m(x,t)$ and consider the following system
of equations
\begin{equation}
\begin{aligned}
&u_t + uu_x + \rho_x = h'(x),\\
&m_x = \rho,\\
&m_t = -u\rho,
\end{aligned}
\label{SWLF}
\end{equation}
where $\rho=\eta+h(x)$.

From the second and third equations it follows that the variable $m$ is
Lagrangian's variable due to the relation
$$
\frac{dm}{dt}=m_t + u m_x = 0.
$$
One can get the equation of one-dimensional shallow water in Lagrange's
variables by choosing $m$ and $t$ as independent
variables~\cite{Chernyi}
\begin{equation}
x_{tt} - \dfrac{x_{mm}}{x_{m}^3} = h'(x).
\label{LE}
\end{equation}

One-to-one correspondence between the system of equations~\eqref{SWLF}
and the equation~\eqref{LE} is given by the relations
\begin{equation}
\begin{aligned}
u = x_t,\qquad \rho = \dfrac{1}{x_m}.
\end{aligned}
\label{Map}
\end{equation}

Note that the system of equations~\eqref{SWLF} is intermediate system
between system of equations~\eqref{SW} and the equation~\eqref{LE}. The
system of equations~\eqref{SWLF} is a covering system~\cite{VinKr} for
the system of equations~\eqref{SW}.

\section{Conservation Laws of the equation in\\ Lagrange's variables}
\label{ConsL}

Under the conservation laws of the system of equations~\eqref{SWLF}
we understand divergent forms for which the solutions of the system
equations~\eqref{SWLF} satisfy the relation
\begin{equation}
D_x(P) + D_t(Q) = 0.
\label{Div}
\end{equation}
Here $P$,  $Q$ are functions of independent and dependent variables and
their derivatives;
$$
D_x=\frac{\partial}{\partial x}+m_x\frac{\partial}{\partial m}+
u_x\frac{\partial}{\partial u}+
\rho_x\frac{\partial}{\partial \rho}+\dots, \quad
D_t=\frac{\partial}{\partial t}+m_t\frac{\partial}{\partial m}+
u_t\frac{\partial}{\partial u}+
\rho_t\frac{\partial}{\partial \rho}+\dots
$$
are total derivatives in variables $x$ and $t$. Conservation laws for
which the equality~\eqref{Div} is satisfied everywhere, we will call
trivial conservation laws. The maximum order of derivatives included in
the functions $P$ and $Q$ will be called the order of the conservation
law. The conservation laws of the zero order will be called
hydrodynamic. The conservation laws of the equation~\eqref{LE} are
defined similarly.

We recall that conservation laws in divergent form~\eqref{Div} are
equivalent to differential 1-forms~\cite{VinKr}
\begin{equation*}
Q\, dx - P\, dt,
\end{equation*}
which are closed on solutions of the system~\eqref{SWLF}.

\textbf{\label{prop1}Proposition~1.} According to the
relations~\eqref{Map}, hydrodynamic conservation law of the system of
equations~\eqref{SWLF} with functions $P$, \ $Q$ defines the
first-order conservation law of the equation~\eqref{LE} with functions
\ $\widetilde{P}=P - x_t\, Q$, \ $\widetilde{Q}=x_mQ$. The opposite is
true.

\noindent
\textbf{Proof.}\, Denote the total derivatives in the variables $m$,
$t$ as $\widetilde{D}_m$ and $\widetilde{D}_t$, and their restrictions
on the equation~\eqref{LE} as $\overline{D}_x$ and $\overline{D}_t$. By
the relations~\eqref{Map}, these derivatives are related in the
following way
\begin{equation*}
\overline{D}_x = \dfrac{1}{x_m}\widehat{D}_m, \qquad
\overline{D}_t = \widehat{D}_t - \dfrac{x_t}{x_m}\widehat{D}_m.
\end{equation*}
Then
\begin{equation*}
\begin{aligned}
\overline{D}_x(P) + \overline{D}_t(Q) &= \dfrac{1}{x_m}\widehat{D}_m(P) +
\widehat{D}_t(Q) - \dfrac{x_t}{x_m}\widehat{D}_m(Q) ={}\\
&= \dfrac{1}{x_m}\Bigl(\widehat{D}_m(P - x_tQ) + \widehat{D}_t(x_mQ)\Bigr).
\end{aligned}
\end{equation*}
This implies the validity of the proposition being proved.

Also true the proposition

\textbf{Proposition~2.} If the functions $P$ and $Q$ in a conservation
law of the system of equations~\eqref{SWLF} are independent of $m$,
then they determine the conservation law of the system of
equations~\eqref{SW}. All conservation laws of the system of
equations~\eqref{SW}, except the conservation law
\begin{equation*}
D_t(\eta) + D_x((\eta+h(x))u),
\end{equation*}
are obtained from the conservation laws of the system of
equations~\eqref{SWLF}.

Note that finding of the hydrodynamic conservation laws of the system
of equations~\eqref{SWLF} is easier than finding of the first-order
conservation laws of the equation~\eqref{LE}.

Relation~\eqref{Div} on solutions of the system of equations~\eqref{SWLF}
takes the form
\begin{equation*}
P_x + u_xP_u + \rho_xP_{\rho} + \rho P_m+Q_t +
(h'(x) - \rho_x - uu_x)Q_u - (u\rho_x + u_x\rho)Q_{\rho} - u\rho Q_m \equiv 0.
\end{equation*}
Equating to zero the coefficients of the derivatives $u_x$ and
$\rho_x$, we obtain the following overdetermined system of linear
equations
\begin{equation}
\begin{aligned}
P_u =&\, \rho Q_{\rho} + uQ_u,\\
P_{\rho} =&\, u Q_{\rho} + Q_u,\\
P_{x} =&\, -\rho P_m + u\rho Q_m - h'Q_u - Q_t.
\end{aligned}
\label{ODS}
\end{equation}
The overdefined system of equations~\eqref{ODS} was investigated on
compatibility.

According to the proposition~\!1, solutions of the system of
equations~\eqref{ODS} can be compared to conservation laws of the
equation~\eqref{LE}. Below we provide functions $\widetilde{P}$ and
$\widetilde{Q}$, which determine the basis of first-order conservation
laws $\widetilde{D}_m(\widetilde{P}) + \widetilde{D}_t(\widetilde{Q})$
of the equation~\eqref{LE} modulo additive trivial conservation laws
for all possible bottom profiles~$h(x)$.

\noindent
\textbf{\mathversion{bold}1. $h=h(x)$ is arbitrary function.} For any
bottom profile $h(x)$, the equation~\eqref{LE} has conservation laws
with functions
\begin{equation*}
\begin{aligned}
&\widetilde{P}_1 = -\dfrac{x_t^2}{2} + \dfrac{1}{x_m} - h(x),\\
&\widetilde{P}_2 = x_t\Bigl(\dfrac{1}{x_m^2} - h^2(x)\Bigr),
\end{aligned}
\qquad
\begin{aligned}
&\widetilde{Q}_1 = x_tx_m,\\
&\widetilde{Q}_2 = x_t^2 + x_m\Bigl(\dfrac{1}{x_m} - h(x)\Bigr)^2.
\end{aligned}
\end{equation*}

\vspace{-0.5ex}
\noindent
\textbf{\mathversion{bold}2. $h=a_1x+a_2$.} In this case
additional conservation laws of the equation~\eqref{LE} correspond to
functions
\begin{equation*}
\begin{aligned}
&\widetilde{P}_3 = (a_1a_2t^2 - 2a_2x)r + ts^2 - a_2^2 t, \\
&\widetilde{Q}_3 = 2tr + \dfrac{2a_2x - a_1a_2t^2}{s} + a_1t^2 - 2x,\\
&\widetilde{P}_4 = 2mr^3 + 24tr^2s^2 - (18x - 9a_1t^2)rs^2 - 12mrs + 16ts^3,\\
&\widetilde{Q}_4 = 16tr(r^2 + 3s) + (9a_1t^2 - 18x)(r^2 + s) -
\dfrac{6mr^2}{s} - 12 m\ln s,\\
&\widetilde{P}_5 = 10trs^2 + 2mr^2 + (3a_1t^2 - 6x)s^2 - 4ms,\\
&\widetilde{Q}_5 = 10tr^2 + 10ts + (6a_1t^2 - 12x)r - \dfrac{4mr}{s},\\
&\widetilde{P}_{\infty} = p(r, s) - r q(r, s),\qquad
\widetilde{Q}_{\infty} = \dfrac{q(r,s)}{s},
\end{aligned}
\end{equation*}
where $r=x_t-a_1t$, $s=1/x_m$;\, functions $p(r, s)$,\, $q(r, s)$ are
arbitrary solutions of the system of equations
\begin{equation*}
p_r = s q_{s} + rq_r\,, \qquad p_{s} = rq_{s} + q_r\,.
\end{equation*}

\vspace{-0.5ex}

\noindent
\textbf{\mathversion{bold}3.1. $h=a_1x^2/2 + a_2x + a_3$, \ $a_1>0$.}
In this case additional conservation laws of the equation~\eqref{LE}
correspond to functions
\begin{equation*}
\begin{aligned}
&\widetilde{P}_3 = e^{-t\sqrt{a_1}}\Bigl(x_thh' +
\dfrac{\sqrt{a_1}}{2}\Bigl(\dfrac{1}{x_m^2} - h^2\Bigr)\Bigr),\\
&\widetilde{Q}_3 = e^{-t\sqrt{a_1}}(\sqrt{a_1}x_t + (1 - x_mh)h'),\\
&\widetilde{P}_4 = e^{t\sqrt{a_1}}\Bigl(x_thh' -
\dfrac{\sqrt{a_1}}{2}\Bigl(\dfrac{1}{x_m^2} - h^2\Bigr)\Bigr),\\
&\widetilde{Q}_4 = e^{t\sqrt{a_1}}(-\sqrt{a_1}x_t + (1 - x_mh)h').
\end{aligned}
\end{equation*}

\vspace{-0.5ex}

\noindent
\textbf{\mathversion{bold}3.2. $h=a_1x^2/2 + a_2x + a_3$, $a_1<0$.} In
this case additional conservation laws of the equation~\eqref{LE}
correspond to functions
\begin{equation*}
\begin{aligned}
&\widetilde{P}_3 = \cos(t\sqrt{-a_1}\,)x_thh' +
\dfrac{\sqrt{-a_1}}{2}\sin(t\sqrt{-a_1}\,)\Bigl(\dfrac{1}{x_m^2} - h^2\Bigr),\\
&\widetilde{Q}_3 = \sin(t\sqrt{-a_1}\,)\sqrt{-a_1}x_t +
\cos(t\sqrt{-a_1}\,)(1 - x_mh)h',\\
&\widetilde{P}_4 = \sin(t\sqrt{-a_1}\,)x_thh' -
\dfrac{\sqrt{-a_1}}{2}\cos(t\sqrt{-a_1}\,)\Bigl(\dfrac{1}{x_m^2} - h^2\Bigr),\\
&\widetilde{Q}_4 = -\cos(t\sqrt{-a_1}\,)\sqrt{-a_1}x_t +
\sin(t\sqrt{-a_1}\,)(1 - x_mh)h'.
\end{aligned}
\end{equation*}

\vspace{-0.5ex}

\noindent
\textbf{\mathversion{bold}4. $h=a_1(x + a_2)^{-4/3} + a_3$, $a_1\neq
0$, $x>-a_2$.} In this case additional conservation laws of the
equation~\eqref{LE} correspond to functions
\begin{equation*}
\begin{aligned}
&\widetilde{P}_3 = -\dfrac{5t x_t}{x_m^2} - mx_t^2 +
\dfrac{3(x + a_2)}{x_m^2} + 2m\Bigl(\dfrac{1}{x_m} - h\Bigr),\\
&\widetilde{Q}_3 = -5t\Bigl(x_t^2 + \dfrac{1}{x_m}\Bigr) +
6(x + a_2) x_t + (10 h - 8a_3)t + 2 mx_tx_m\,.
\end{aligned}
\end{equation*}

\vspace{-0.5ex}

The conservation laws for cases 2\,--\,4 are additional conservation
laws to the conservation laws of the general case~1.

\section{Comparison of conservation laws in Euler's and Lagrangian's variables}

Conservation laws of the equation~\eqref{LE} that do not correspond to
the conservation laws of the system of equations~\eqref{SW} are
obtained from the conservation laws of the system of
equations~\eqref{SWLF}, which are depend on the Lagrangian variable
$m$. The results of the section~\ref{ConsL} show that such a
conservation laws are exist only in two cases. In the case of
$h=a_1x+a_2$ conservation laws, which are defined by the functions
$(\widetilde{P}_4, \widetilde{Q}_4)$ and $(\widetilde{P}_5,
\widetilde{Q}_5)$, are not correspond to the conservation laws of the
system of equations~\eqref{SW}; \ in the case $h=a_1(x + a_2)^{-4/3} +
a_3$ \ ($a_1\neq 0$, \ $x>-a_2$) conservation laws, which are defined
by the functions $(\widetilde{P}_3, \widetilde{Q}_3)$, also do not
comply with the conservation laws of the system of
equations~\eqref{SW}. The system of shallow water equations in Eulerian
variables~\eqref{SW} has no additional conservation laws in this
case~\cite{AksDr}. All other first-order conservation laws of the
equation~\eqref{LE} are correspond to the conservation laws of the
system of equations~\eqref{SW}.

\section{Acknowledgments}

The authors wish to express gratitude to V.A.~Dorodnitsyn for
constructive discussions and helpful remarks.

The work was supported by the Russian Science Foundation under grant
18-11-00238.

\end{document}